\documentclass[sigconf,table]{acmart}

\usepackage{booktabs} 
\usepackage{tabularx} 
\usepackage[english]{babel}
\usepackage{mathrsfs}

\usepackage{multirow}
\usepackage{multicol}
\usepackage{array}
\usepackage{arydshln}

\usepackage{amsmath}
\usepackage{amssymb}
\usepackage{booktabs}
\usepackage[inline]{enumitem}
\usepackage{siunitx}
\usepackage{stackengine}
\usepackage{extarrows}

\usepackage{caption}
\usepackage{subcaption}

\usepackage{footnote}
\makesavenoteenv{tabular}
\makesavenoteenv{table}

\usepackage{xspace} 

\newcolumntype{L}[1]{>{\raggedright\arraybackslash}p{#1}}
\newcolumntype{C}[1]{>{\centering\arraybackslash}p{#1}}
\newcolumntype{R}[1]{>{\raggedleft\arraybackslash}p{#1}}

\copyrightyear{2019} 
\acmYear{2019} 
\acmConference[Pre-Print]{}{Microsoft}{AI \& Research}

\author{Hamed Zamani and Nick Craswell}
\affiliation{%
  \institution{Microsoft AI \& Research}
    \country{United States}
}
\email{{hazamani, nickcr}@microsoft.com}

\begin{document}

\title{Macaw: An Extensible Conversational Information Seeking Platform}

\begin{abstract}
Conversational information seeking (CIS) has been recognized as a major emerging research area in information retrieval. Such research will require data and tools, to allow the implementation and study of conversational systems. This paper introduces \emph{Macaw}, an open-source framework with a modular architecture for CIS research. Macaw supports multi-turn, multi-modal, and mixed-initiative interactions, and enables research for tasks such as document retrieval, question answering, recommendation, and structured data exploration. It has a modular design to encourage the study of new CIS algorithms, which can be evaluated in batch mode. It can also integrate with a user interface, which allows user studies and data collection in an interactive mode, where the back end can be fully algorithmic or a wizard of oz setup. Macaw is distributed under the MIT License.\footnote{Macaw is available at \url{https://github.com/microsoft/macaw}.} 
\end{abstract}
\keywords{Conversational Search; Conversational Question Answering; Conversational Recommendation; Conversational Natural Language Interface; Open-Source Library}
\maketitle

\section{Introduction}
\label{sec:intro}
The rapid growth in speech and small screen interfaces, particularly on mobile devices, has significantly influenced the way users interact with intelligent systems to satisfy their information needs. The growing interest in personal digital assistants, such as Amazon Alexa, Apple Siri, Google
Assistant, and Microsoft Cortana, demonstrates the willingness of users to employ conversational interactions~\cite{Radlinski:2017}. As a result, conversational information seeking (CIS) has been recognized as a major emerging research area in the Third Strategic Workshop on Information Retrieval (SWIRL 2018)~\cite{Culpepper:2018}.\footnote{\url{https://sites.google.com/view/swirl3/}} 


Research progress in CIS relies on the availability of resources to the community. There have been recent efforts on providing data for various CIS tasks, such as the TREC 2019 Conversational Assistance Track (CAsT),\footnote{\url{http://www.treccast.ai/}} MISC~\cite{Thomas:2017}, Qulac~\cite{Aliannejadi:2019}, CoQA~\cite{Reddy:2019}, QuAC~\cite{Choi:2018}, SCS~\cite{Trippas:2019}, and CCPE-M~\cite{Radlinski:2019}. In addition, \citet{Dalton:2018} have implemented a demonstration for conversational movie recommendation based on Google's DialogFlow. Despite all of these resources, the community still feels the lack of a suitable platform for developing CIS systems. We believe that providing such platform will speed up the progress in conversational information seeking research. Therefore, we developed a general framework for supporting CIS research. The framework is called \emph{Macaw}. This paper describes the high-level architecture of Macaw, the supported functionality, and our future vision. Researchers working on various CIS tasks should be able to take advantage of Macaw in their projects.

\begin{figure}[t]
\centering
\begin{subfigure}{.23\textwidth}
  \centering
  \includegraphics[width=\linewidth]{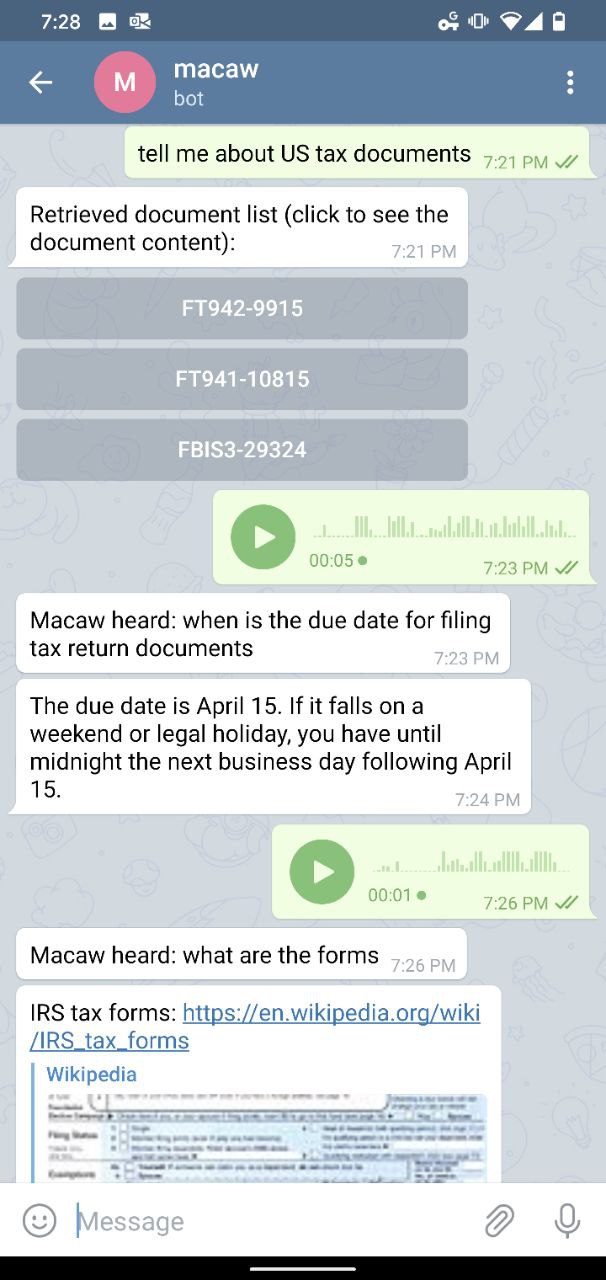}
  \caption{Multi-modal interactions.}
  \label{fig:sub1}
\end{subfigure}%
\begin{subfigure}{.1\textwidth}
\end{subfigure}
\begin{subfigure}{.23\textwidth}
  \centering
  \includegraphics[width=\linewidth]{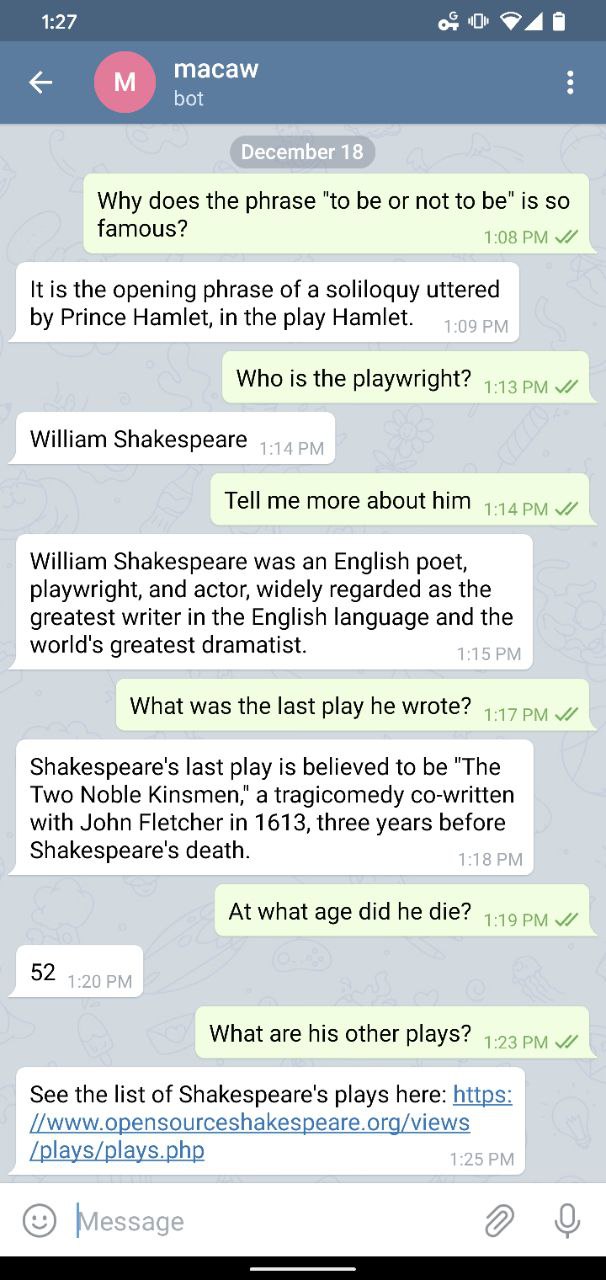}
  \caption{Multi-turn interactions.}
  \label{fig:sub2}
\end{subfigure}
\caption{Example screenshots of the Macaw interface on mobile devices using Telegram bots. Macaw supports multi-modal and multi-turn interactions.}
\label{fig:screenshots}
\end{figure}

Macaw is designed based on a \textbf{modular architecture} to support different information seeking tasks, including conversational search, conversational question answering, conversational recommendation, and conversational natural language interface to structured and semi-structured data. Each interaction in Macaw (from both user and system) is a \emph{Message} object, thus a conversation is a list of Messages. Macaw consists of multiple actions, each action is a module that can satisfy the information needs of users for some requests. For example, search and question answering can be two actions in Macaw. Even multiple search algorithms can be also seen as multiple actions. Each action can produce multiple outputs (e.g., multiple retrieved documents). For every user interaction, Macaw runs all actions in parallel. The actions' outputs produced within a predefined time interval (i.e., an interaction timeout constant) are then post-processed. Macaw can choose one or combine multiple of these outputs and prepare an output Message object as the user's response.

\begin{figure*}[t]
    \centering
    \includegraphics[trim={0 3cm 0 3cm},clip,width=.8\textwidth]{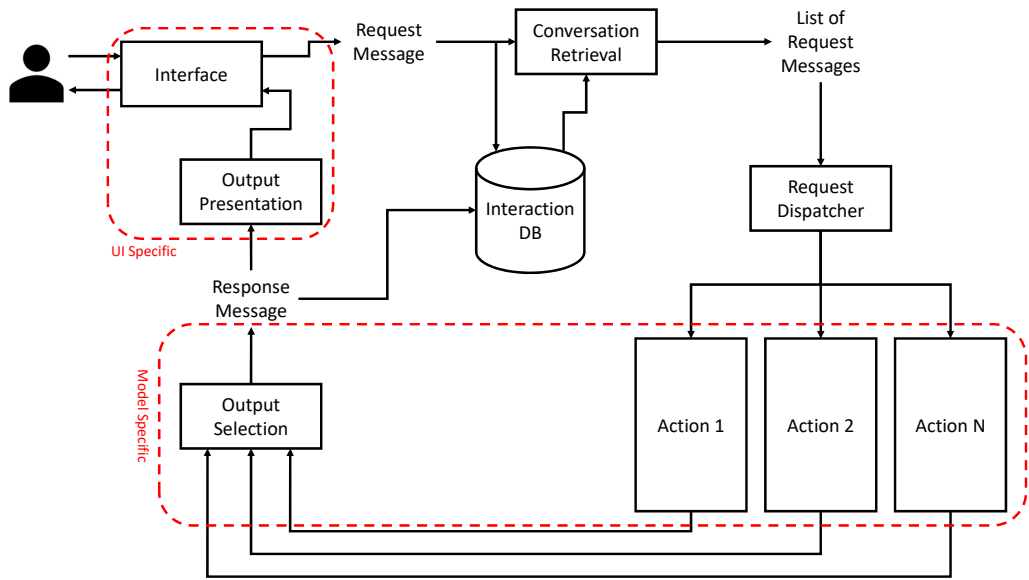}
    \caption{The high-level architecture of Macaw for developing conversation information seeking systems.}
    \label{fig:arch}
\end{figure*}

The modular design of Macaw makes it relatively easy to configure a different user interface or add a new one. The current implementation of Macaw supports a command line interface as well as mobile, desktop, and web apps. In more detail, Macaw's interface can be a \emph{Telegram} bot, which supports a wide range of devices and operating systems (see \figurename~\ref{fig:screenshots}). This allows Macaw to support \textbf{multi-modal} interactions, such as text, speech, image, click, etc. A number of APIs for automatic speech recognition and generation have been employed to support speech interactions. Note that the Macaw's architecture and implementation allows \textbf{mixed-initiative} interactions.

The research community can benefit from Macaw for the following purposes:
\begin{itemize}[leftmargin=*]
    \item Developing algorithms, tools, and techniques for CIS.
    \item Studying user interactions with CIS systems.
    \item Performing CIS studies based on an intermediary person and wizard of oz.
    \item Preparing quick demonstration for a developed CIS model.
\end{itemize}

\section{Macaw Architecture}
\label{sec:arch}
Macaw has a modular design, with the goal of making it easy to configure and add new modules such as a different user interface or different retrieval module. The overall setup also follows a Model-View-Controller (MVC) like architecture. The design decisions have been made to smooth the Macaw's adoptions and extensions. Macaw is implemented in Python, thus machine learning models implemented using PyTorch,\footnote{\url{https://pytorch.org/}} Scikit-learn,\footnote{\url{https://scikit-learn.org/}} or TensorFlow\footnote{\url{http://tensorflow.org/}} can be easily integrated into Macaw. The high-level overview of Macaw is depicted in \figurename~\ref{fig:arch}. The user interacts with the interface and the interface produces a Message object from the current interaction of user. The interaction can be in multi-modal form, such as text, speech, image, and click. Macaw stores all interactions in an ``Interaction Database''. For every interaction, Macaw looks for most recent user-system interactions (including the system's responses) to create a list of Messages, called the conversation list. It is then dispatched to multiple information seeking (and related) actions. The actions run in parallel, and each should respond within a pre-defined time interval. The output selection component selects from (or potentially combines) the outputs generated by different actions and creates a Message object as the system's response. This message is logged into the interaction database and is sent to the interface to be presented to the user. Again, the response message can be multi-modal and include text, speech, link, list of options, etc.

\begin{figure*}[t]
    \centering
    \includegraphics[trim={0 3cm 0 3cm},clip,width=.8\textwidth]{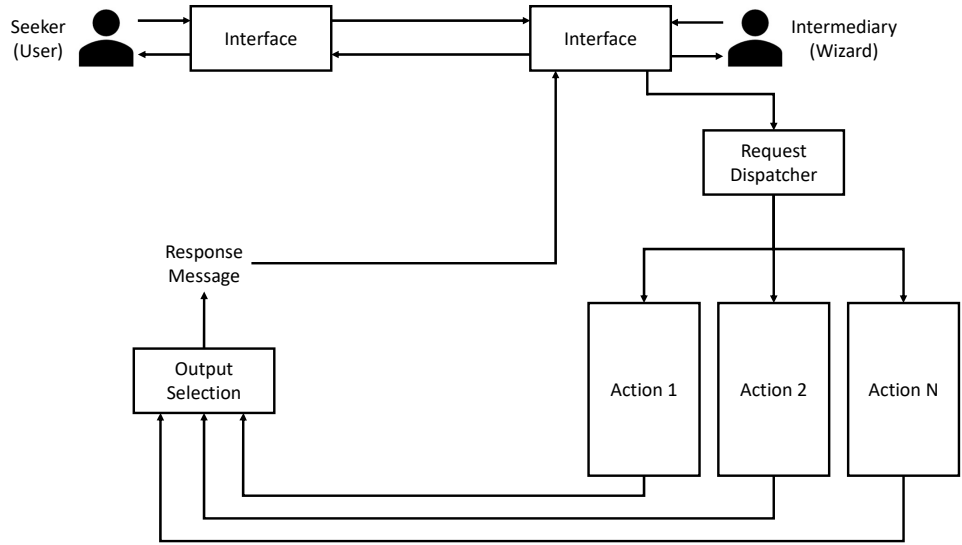}
    \caption{The high-level architecture of Macaw for user studies. In this architecture, user interacts with a human intermediary who is an expert of the system and can interact with the system to address the user's information need.}
    \label{fig:woz}
\end{figure*}

Macaw also supports Wizard of Oz studies or intermediary-based information seeking studies. The architecture of Macaw for such setup is presented in \figurename~\ref{fig:woz}. As shown in the figure, the seeker interacts with a real conversational interface that supports multi-modal and mixed-initiative interactions in multiple devices. The intermediary (or the wizard) receives the seeker's message and performs different information seeking actions with Macaw. All seeker-intermediary and intermediary-system interactions will be logged for further analysis. This setup can simulate an ideal CIS system and thus is useful for collecting high-quality data from real users for CIS research.

\begin{figure}[t]
    \centering
    \includegraphics[trim={5cm 6cm 5cm 5cm},clip,width=\linewidth]{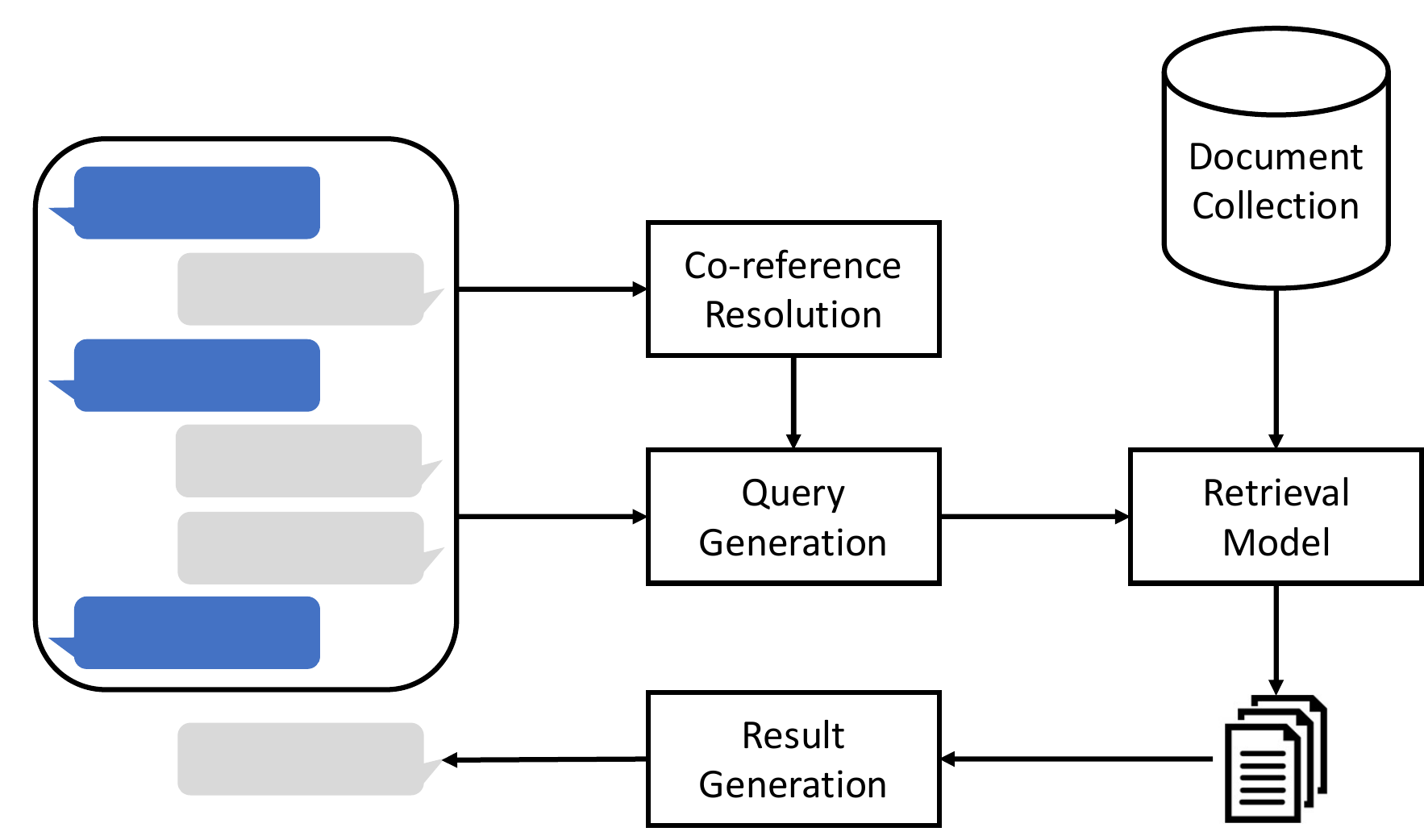}
    \caption{The overview of retrieval and question answering in Macaw.}
    \label{fig:retrieval}
\end{figure}

\section{Retrieval and Question Answering in Macaw}
\label{sec:retrieval}
The overview of retrieval and question answering actions in Macaw is shown in \figurename~\ref{fig:retrieval}. These actions consist of the following components:
\begin{itemize}[leftmargin=*]
    \item \textbf{Co-Reference Resolution:} To support multi-turn interactions, it is sometimes necessary to use co-reference resolution techniques for effective retrieval. In Macaw, we identify all the co-references from the last request of user to the conversation history. The same co-reference resolution outputs can be used for different query generation components. This can be a generic or action-specific component.
    
    \item \textbf{Query Generation:} This component generates a query based on the past user-system interactions. The query generation component may take advantage of co-reference resolution for query expansion or re-writing. 
    
    \item \textbf{Retrieval Model:} This is the core ranking component that retrieves documents or passages from a large collection. Macaw can retrieve documents from an arbitrary document collection using the Indri python interface~\cite{Strohman:2005,VanGysel:2017}.\footnote{Indri~\cite{Strohman:2005} is an open-source search engine originally implemented to support language models in information retrieval as part of the Lemur Project (\url{http://lemurproject.org/}). It features a wide range of retrieval models. For more information visit \url{http://lemurproject.org/indri.php}.} We also provide the support for web search using the Bing Web Search API.\footnote{\url{https://azure.microsoft.com/en-us/services/cognitive-services/bing-web-search-api/}} Macaw also allows multi-stage document re-ranking.
    
    \item \textbf{Result Generation:} The retrieved documents can be too long to be presented using some interfaces. Result generation is basically a post-processing step ran on the retrieved result list. In case of question answering, it can employ answer selection or generation techniques, such as machine reading comprehension models. For example, Macaw features the DrQA model~\cite{Chen:2017} for question answering. 
\end{itemize}

These components are implemented in a generic form, so researchers can easily replace them with their own favorite algorithms.

\section{User Interfaces}
\label{sec:interface}
We have implemented the following interfaces for Macaw:
\begin{itemize}[leftmargin=*]
    \item \textbf{File IO:} This interface is designed for \emph{experimental} purposes, such as evaluating the performance of a conversational search technique on a dataset with multiple queries. This is not an interactive interface.
    
    \item \textbf{Standard IO:} This interactive command line interface is designed for \emph{development} purposes to interact with the system, see the logs, and debug or improve the system.
    
    \item \textbf{Telegram:} This interactive interface is designed for \emph{interaction with real users} (see \figurename~\ref{fig:screenshots}). Telegram\footnote{\url{https://telegram.org/}} is a popular instant messaging service whose client-side code is open-source. We have implemented a Telegram bot that can be used with different devices (personal computers, tablets, and mobile phones) and different operating systems (Android, iOS, Linux, Mac OS, and Windows). This interface allows multi-modal interactions (text, speech, click, image). It can be also used for speech-only interactions. For speech recognition and generation, Macaw relies on online APIs, e.g., the services provided by Google Cloud and Microsoft Azure. In addition, there exist multiple popular groups and channels in Telegram, which allows further integration of social networks with conversational systems. For example, see the Naseri and Zamani's study on news popularity in Telegram~\cite{Naseri:2019}.
\end{itemize}

Similar to the other modules, one can easily extend Macaw using other appropriate user interfaces.

\section{Limitations and Future Work}
\label{sec:limits}
The current implementation of Macaw lacks the following actions. We intend to incrementally improve Macaw by supporting more actions and even more advanced techniques for the developed actions.

\begin{itemize}[leftmargin=*]
    \item \textbf{Clarification and Preference Elicitation:} Asking clarifying questions has been recently recognized as a necessary component in a conversational system~\cite{Aliannejadi:2019,Radlinski:2019}. The authors are not aware of a published solution for generating clarifying questions using public resources. Therefore, Macaw does not currently support clarification.
    
    \item \textbf{Explanation:} Despite its importance, result list explanation is also a relatively less explored topic. We intend to extend Macaw with result list explanation as soon as we find a stable and mature solution.
    
    \item \textbf{Recommendation:} In our first release, we focus on conversational search and question answering tasks. We intend to provide support for conversational recommendation, e.g., \cite{Li:2018,Sun:2018,Zhang:2018}, and joint search and recommendation, e.g., \cite{Zamani:2018:DESIRES,Zamani:2020:WSDM}, in the future.
    
    \item \textbf{Natural Language Interface:} Macaw can potentially support access to structured data, such as knowledge graph. We would like to ease conversational natural language interface to structured and semi-structured data in our future releases.
\end{itemize}

\section{Contribution}
\label{sec:contrib}
Macaw is distributed under the MIT License. We welcome contributions and suggestions. Most contributions require you to agree to a Contributor License Agreement (CLA) declaring that you have the right to, and actually do, grant us the rights to use your contribution. For details, visit \url{https://cla.opensource.microsoft.com}. This project has adopted the Microsoft Open Source Code of Conduct.

When you submit a pull request, a CLA bot will automatically determine whether you need to provide a CLA and decorate the PR appropriately (e.g., status check, comment). Simply follow the instructions provided by the bot. You will only need to do this once across all repos using our CLA.

\section{Conclusions}
\label{sec:conc}
This paper described Macaw, an open-source platform for conversational information seeking research. Macaw supports multi-turn, multi-modal, and mixed-initiative interactions. It was designed based on a modular architecture that allows further improvements and extensions. Researchers can benefit from Macaw for developing algorithms and techniques for conversational information seeking research, for user studies with different interfaces, for data collection from real users, and for preparing a demonstration of a CIS model.

\section{Acknowledgements}
The authors wish to thank Ahmed Hassan Awadallah, Krisztian Balog, and Arjen P. de Vries for their invaluable feedback.



\end{document}